\documentclass[useAMS,usenatbib]{mn2e}

\usepackage[english]{babel}
\usepackage{fancyhdr}
\usepackage{amsfonts}
\usepackage{amsmath}
\usepackage{amssymb}
\usepackage{multicol}
\usepackage{layout}
\usepackage{psfig}
\usepackage{graphicx}
\usepackage{subfigure}
\usepackage[colorlinks,linkcolor=blue,citecolor=blue,urlcolor=blue,pdftitle={Shear-flexion cross-talk in weak-lensing measurements},pdfauthor={Viola, Melchior & Bartelmann}]{hyperref}

\usepackage{times}
\usepackage{natbib}

\newif\ifAMStwofonts
\AMStwofontstrue

\renewcommand{\vec}[1]{\bmath{#1}}
\newcommand{\be}{\begin{equation}}
\newcommand{\ee}{\end{equation}}
\newcommand{\ba}{\begin{eqnarray}}
\newcommand{\ea}{\end{eqnarray}}
\newcommand{\brr}{\begin{array}}
\newcommand{\err}{\end{array}}
\newcommand{\bc}{\begin{center}}
\newcommand{\ec}{\end{center}}

\newcommand{\mincir}{\raise
  -2.truept\hbox{\rlap{\hbox{$\sim$}}\raise5.truept \hbox{$<$}\ }}
\newcommand{\magcir}{\raise
  -2.truept\hbox{\rlap{\hbox{$\sim$}}\raise5.truept \hbox{$>$}\ }}
\newcommand{\siml}{\raise
  -2.truept\hbox{\rlap{\hbox{$\sim$}}\raise5.truept \hbox{$<$}\ }}
\newcommand{\simg}{\raise
  -2.truept\hbox{\rlap{\hbox{$\sim$}}\raise5.truept \hbox{$>$}\ }}

\newcommand {\apgt} {\ {\raise-.5ex\hbox{$\buildrel>\over\sim$}}\ }
\newcommand {\aplt} {\ {\raise-.5ex\hbox{$\buildrel<\over\sim$}}\ }

\title{Shear-flexion cross-talk in weak-lensing measurements}  \author[Viola, Melchior \& Bartelmann] {M. Viola$^{1,2}$, P. Melchior$^{3,4}$, M. Bartelmann$^{1}$ \\~\\
  $^1$ Zentrum f\"ur Astronomie, ITA, Universit\"at Heidelberg, Albert-Ueberle-Str.2, 69120 Heidelberg, Germany \\
  $^2$ Institute for Astronomy, University of Edinburgh, Royal Observatory, Blackford Hill, Edinburgh, EH93HJ, U.K.\\
  $^{3}$Center for Cosmology and Astro-Particle Physics, The Ohio State University, 191 W. Woodruff Ave., Columbus, Ohio 43210, USA\\
$^{4}$Department of Physics, The Ohio State University, 191 W. Woodruff Ave., Columbus, Ohio 43210, USA\\~\\
  E-mail: mv@roe.ac.uk}
\date{Accepted 2011 September 21.  Received 2011 September 12; in original form 2011 July 20}
\pagerange{\pageref{firstpage}--\pageref{lastpage}}

\begin{document}
\label{firstpage}
\maketitle

\begin{abstract}
Gravitational flexion, caused by derivatives of the gravitational tidal field, is potentially important for the analysis of the dark-matter distribution in gravitational lenses, such as galaxy clusters or the dark-matter haloes of galaxies. Flexion estimates rely on measurements of galaxy-shape distortions with spin-1 and spin-3 symmetry. We show in this paper that and how such distortions are generally caused not only by the flexion itself, but also by coupling terms of the form (shear $\times$ flexion), which have hitherto been neglected. Similar coupling terms occur between intrinsic galaxy ellipticities and the flexion. We show, by means of numerical tests, that neglecting these terms can introduce biases of up to 85\% on the $F$ flexion and 150\% on the $G$ flexion for galaxies with an intrinsic ellipticity dispersion of $\sigma_{\epsilon}=0.3$. In general, this bias depends on the strength of the lensing fields, the ellipticity dispersion, and the concentration of the lensed galaxies.
We derive a new set of equations relating the measured spin-1 and spin-3 distortions to the lensing fields up to first order in the shear, the flexion, the product of shear and flexion, and the morphological properties of the galaxy sample. We show that this new description is accurate with a bias $\leq 7\%$ (spin-1 distortion) and $\leq 3\%$  (spin-3 distortion) even close to points where the flexion approach breaks down due to merging of multiple images.
We propose an explanation why a spin-3 signal could not be measured yet and comment on the potential difficulties in using a model-fitting approach to measure the flexion signal.
\end{abstract}


\section{Introduction}

In the last years weak gravitational lensing became a standard technique to infer the mass distribution of galaxy clusters. In contrast to X-ray or galaxy dynamics, which typically assume hydrostatic or virial equilibrium and spherical mass distributions, gravitational lensing allows to directly probe the projected gravitational potential: 
Light from distant galaxies gets deflected by the cluster potential and the galaxy ellipticities are modified by the gravitational tidal field of the intervening matter. Hence measurements of the galaxy ellipticities allow to directly trace the curvature of the gravitational potential.

Beyond that, higher-order distortions of background galaxies, like curvature or distortions with three-fold rotational symmetry, can be related to higher-order derivatives of the gravitational potential, thereby providing information on the variation of the potential at smaller scales. 
By expanding the lens equation to second order, one can identify two combinations of third-order derivatives of the potential, called $F$ and $G$ flexions. These two fields can be related to a skewness in the light distribution and distortions with three-fold rotational symmetry in the shape of galaxies, respectively \citep{Goldberg05,Bacon06,Irwin07}. Reliable measurements of these higher-order distortions can increase the accuracy with which the inner profile of galaxy clusters can be constrained or can support the detection of substructures \citep{Bacon06}, whose presence in galaxy clusters is a firm prediction of the Cold-Dark Matter cosmogony \citep{White78}. Moreover, at galactic scales the incorporation of flexion constraints significantly improves the recovery of halo ellipticity in comparison to shear-only measurements \citep{Hawken09}.

Two main approaches to measure these higher-order distortions in the galaxy shape have been proposed so far: shapelets \citep{Goldberg05,Velander11} and HOLICs \citep{Okura07,Okura08}. Both assume that the measured  distortions can simply be related to the $F$- and $G$-flexions, respectively. This is true only if the shear can be considered negligible and galaxies are intrinsically circular. We will show that a considerable error on the inferred lens power is introduced if any of these assumptions is violated. Then, coupling terms of the form (shear $\times$ flexion) or (ellipticity $\times$ flexion) create additional distortions, which are indistinguishable from the pure flexion-induced distortions and thus interfere with the interpretation of the lensing signal.

In \autoref{sec:flexion_moments} we derive a new set of equations, relating the measured distortion to the lensing fields, including these coupling terms. In \autoref{sec:morphology} we discuss the physical meaning of the newly identified terms. In \autoref{sec:new_estimators} we derive a simplified set of equations, applicable to lenses with non-vanishing shear, which also consider the responsivity of the higher-order distortions to shear. 
These new equations are more general than the equation originally presented by \citet{Okura08} and constitute a practical, yet accurate approximation to the equations presented by \citet{Schneider08}.

By means of numerical tests, we show in \autoref{sec:tests} that our equations are sufficient to reliably describe the lensing-induced change of galaxy shapes up to the point where the very concept of flexion estimates from individual galaxies breaks down due to merging of multiple images of the same source \citep{Schneider08}. In \autoref{sec:discussion} we propose an explanation why the $G$-flexion-related spin-3 field has not been measured yet \citep{Goldberg05,Leonard07, Okura08, Leonard11} despite the fact that the $G$-flexion extends to larger cluster-centric distances than the $F$-flexion, and we comment of the conceptual difficulties of using a model-fitting approach to measure the flexion signal. We conclude in \autoref{sec:conclusions}. 

\section{Flexion formalism}
\label{sec:formalism}

This section summarises the basic weak-lensing concepts used throughout this work. For a complete overview we refer to \cite{Bartelmann01}. 
An isolated lens with surface mass density $\Sigma(\vec{\theta})$ has the lensing potential
\begin{equation}
\Psi(\vec{\theta})=\frac{4G}{c^2}\frac{D_{l}D_{s}}{D_{ls}}\int d^2\theta^{\prime}\Sigma(\vec{\theta}^{\prime})\ln |\vec{\theta}-\vec{\theta}^{\prime}|, \label{eq:potential}
\end{equation}
where $G$ denotes the gravitational constant, $c$ the speed of light, and $D_{l, s, ls}$ are the angular-diameter distances between the observer and the lens, the observer and the source, and the lens and the source, respectively.
In the approximation of a single lens plane,\footnote{This approximation provides a very accurate description of most lensing phenomena, including flexion \citep{Schaefer11}.} light rays are deflected by the angle
\begin{equation}
\vec{\alpha}(\vec{\theta})=\nabla \Psi(\vec{\theta})\;,\label{eq:deflection}
\end{equation}
which relates the angular positions on the source plane $\vec\beta$ and those on the apparent image plane $\vec\theta$ by the lens equation
\begin{equation}
\vec{\beta}=\vec{\theta}-\vec{\alpha}(\vec{\theta})\;.
\label{eq:lenseq}
\end{equation}

If we assume that the source is much smaller than the scale on which properties of the lens change, we can locally expand the lens equation:
\begin{equation}
\beta_{i}\simeq\theta_{i}-\Psi_{,ij}\theta_{j}-\frac{1}{2}\Psi_{,ijk}\theta_{j}\theta_{k}\;.
\label{eq:linlenseq}
\end{equation}
In many weak lensing applications the expansion of the lens equation can be safely truncated at first order, such that the image distortion is described by the Jacobian matrix
\begin{equation}
\begin{split}
A\equiv\frac{\partial\vec{\beta}}{\partial \vec{\theta}}&=\left(\delta_{ij}-\frac{\partial ^2\Psi(\vec{\theta})}{\partial \theta_{i}\partial \theta_{j}}\right)\\
&=\left(\begin{array}{cc}
1-\kappa -\gamma_1 & -\gamma_2 \\
-\gamma_2 & 1-\kappa + \gamma_1 \end{array} \right)\;, 
\end{split}
\label{eq:distortion}
\end{equation}
where $\kappa$ is the convergence and the $\gamma_i$ are the two components of the shear. If, however, the sources are large or the lens properties change rapidly on small scales (e.g.\ towards the core of a galaxy cluster), second-order terms become important.\footnote{Eventually, even the second-order terms cannot fully describe all lensing effects, for instance the occurrence of giant arcs. In these cases the full \autoref{eq:lenseq} has to be studied.}

At this point, it is convenient to introduce a complex notation for the relevant lensing quantities. 
We start defining a complex gradient operator \citep{Newman62,Bacon06}
\begin{equation}
\partial =\frac{\partial}{\theta_{1}}+i\frac{\partial}{\theta_2}\equiv \partial_{1} + i \partial_{2},
\end{equation}
which in polar coordinates has the form
\begin{equation}
\partial=e^{i\phi}\Bigg(\frac{\partial}{\partial r}+\frac{i}{r}\frac{\partial}{\partial \theta}\Bigg).
\end{equation}
This representation shows clearly that when $\partial$ is applied to a spin-$s$ quantity,\footnote{We say that a quantity has spin $s$ if it is invariant under a rotation of the Cartesian coordinate frame by a rotation angle $\phi = 2\pi/s$ and $s \in [1, 2, ...]$. Vectors are thus spin-1 quantities.} it raises its spin by one. Analogously, $\partial^*$ lowers the spin by one.
Applying this operator one time to the deflection potential (spin-0), we generate the deflection angle
\begin{equation}
\alpha=\partial \Psi,
\end{equation}
a spin-1 field. Applying it twice, we generate the spin-0 convergence field
\begin{equation}
\kappa=\frac{1}{2}\partial \partial^{*}\Psi=\frac{1}{2}(\Psi_{,11}+\Psi_{,22})
\end{equation}
or the spin-2 shear field
\begin{equation}
\gamma=\frac{1}{2}\partial \partial\Psi=\frac{1}{2}[(\Psi_{,11}-\Psi_{,22})+2i\Psi_{,12}],
\end{equation}
while applying it three times generates the so-called \textit{$\mathcal{F}$-flexion} (spin-1)
\begin{equation}
\mathcal{F}=\frac{1}{2}\partial\partial \partial^{*}\Psi=\frac{1}{2}[(\Psi_{,111}+\Psi_{,122})+i(\Psi_{,112}+\Psi_{,222})]
\end{equation}
or the \textit{$\mathcal{G}$-flexion} (spin-3)
\begin{equation}
\mathcal{G}=\frac{1}{2}\partial \partial \partial\Psi=\frac{1}{2}[(\Psi_{,111}-3\Psi_{,122})+i(3\Psi_{,112}-\Psi_{,222})]
\end{equation}
fields. We can now use these new fields to rewrite \autoref{eq:linlenseq} in complex notation:
\begin{equation}
\beta\simeq(1-\kappa)\Bigg(\theta -g \theta^{*}-\frac{1}{4}F^{*}\theta^{2}-\frac{1}{2}F\theta\theta^{*}-\frac{1}{4}G(\theta^{*})^{2}\Bigg),
\label{eq:lenseqfield}
\end{equation}
where the reduced shear and flexions are defined as
\begin{equation}
g\equiv \frac{\gamma}{1-\kappa},\ \  F\equiv \frac{\mathcal{F}}{1-\kappa},\ \  G \equiv \frac{\mathcal{G}}{1-\kappa}\;.
\end{equation}
This equation relates source plane and observed positions in terms of derivative of the lensing potential with uniquely defined spin properties. One can show \citep{Schneider08} that the Jacobian of this transformation is
\begin{equation}
\begin{split}
\det A \simeq\ & (1-\kappa)^{2}\Bigg[1  - gg^{\star} -\theta\Bigg[F^{\star}+\frac{1}{2}\Bigg(g^{\star}F+gG^{\star}\Bigg)\Bigg]- \\
&\theta^{\star}\Bigg[F+\frac{1}{2}\Bigg(g^{\star}G+gF^{\star}\Bigg)\Bigg]\Bigg]+\mathcal{O}(\theta^2)\;.
\label{eq:determinant}
\end{split}
\end{equation}

\section{Measuring flexion from moments}
\label{sec:flexion_moments}

Gravitational flexion is related to the third derivative of the lensing potential (third-order lensing effect) and it manifests itself in the skewed and arclike appearance of lensed galaxies. 
A possible description of these higher-order distortions can be given in terms of moments of the galaxy surface brightness,
\begin{equation}
Q_{ij\ldots k}=\frac{1}{S_{0}}\int d^2 \theta I(\theta)\theta_{i}\theta_{j}\ldots\theta_{k}\;.
\end{equation}
where $S_{0}$ denotes the flux (0-th moment).
For instance, a common ellipticity estimator can be formed from second-order moments,
\begin{equation}
\chi = \frac{a^2-b^2}{a^2+b^2}e^{2i\phi} = \frac{(Q_{11}-Q_{22})+2iQ_{12}}{Q_{11}+Q_{22}}\;,
\label{eq:chi}
\end{equation} 
where $a$ and $b$ are the semi-major and semi-minor axes of the ellipse.
Consequently, it is possible to form combinations of third-order moments such that they exhibit spin-1 or spin-3 properties, which we then seek to relate to the flexion fields. In this work, we follow the notation introduced by \cite{Okura08} to describe these higher-order distortions:

\begin{equation}
\begin{split}
\zeta  \equiv \frac{1}{\xi S_{0}}\int d^2 \theta I(\theta)\theta^2 \theta^{\star} = \frac{(Q_{111}+Q_{122})+i(Q_{112}+Q_{222})}{\xi}\\
\delta \equiv \frac{1}{\xi S_{0}}\int d^2 \theta I(\theta)\theta^3 = \frac{(Q_{111}-3Q_{122})+i(3Q_{112}-Q_{222})}{\xi}\;.
\label{eq:zetaDef} 
\end{split}
\end{equation}
These quantities have spin 1 and spin 3, respectively. The normalisation factor has spin 0 and it is defined as a combination of fourth-order moments,
\begin{equation}
\xi \equiv \frac{1}{S_{0}}\int d^2 \theta I(\theta)(\theta \theta^{\star})^2 = Q_{1111}+2Q_{1122}+Q_{2222}\;,
\end{equation}
\citet{Okura08} derived the lensing transformations of $\zeta$ and $\delta$, considering only terms of first order in the flexion fields. Specifically, terms of the kind \emph{shear $\times$ flexion} were discarded, which is equivalent to assuming a vanishing shear. As we shall show in \autoref{sec:tests} using the example of a realistic NFW halo, flexion will necessarily be measured in regions of the lens where the shear is \emph{not} negligible (cf. \autoref{fig:profile}).
\citet{Schneider08} derived a complete set of transformations of 3rd- and 4th-order moments under lensing in a somewhat different notation, but noticed biases of their flexion estimates that remained unexplained.

We therefore re-derive the transformations of $\zeta$ and $\delta$ under lensing, keeping all terms of first order in the shear, the flexion, and the product of shear and flexion. The calculation proceeds along the following scheme (and can be reviewed by the interested and courageous reader in \autoref{sec:flexion_trafo}):
\begin{enumerate}
\item Compute the transformation of 3rd- and 4th-order moments under lensing according to \autoref{eq:lenseqfield}. This computation is done neglecting the centroid shift induced by lensing.
\item Correct for the lensing induced centroid shift.
\end{enumerate}
The results of these calculations show that at first order in the flexion and shear fields the transformations of $\zeta$ and $\delta$ under lensing are
\begin{subequations}
\label{eq:flexionfull}
\begin{equation}
\begin{split}
\zeta^{\mathrm{s}}=&\frac{1}{[1-4\Re({g^*\eta})](1-\kappa)}\Bigg[\zeta-2g\zeta^*-g^*\delta - 2F^{\star}\eta \\
& -\frac{9}{4}F -\frac{1}{2}G\eta^{\star} -\frac{1}{4} G^{*}\lambda+ 3F^{\star}g+\frac{3}{2}Fg^{\star}\eta + \frac{3}{2}(F^{\star}g^{\star}\lambda)\\
&+\frac{1}{2}Gg\lambda^{\star} +\frac{7}{2} Fg\eta^{\star}+ \mu\Bigg(4F^{\star}\chi+\frac{1}{2}Fg^{\star}\chi+gG^{\star}\chi +3F \\
&+Gg^{\star}-4F^{\star}g -\frac{9}{2}gF\chi^{\star}+\frac{1}{2}G\chi^{\star}-3F\Re(g\chi^{\star})\Bigg)\Bigg]
\label{eq:zetafull}
\end{split}
\end{equation}
and
\begin{equation}
\begin{split}
\delta^{\mathrm{s}}=&\frac{1}{[1-4\Re({g^*\eta})](1-\kappa)} \Bigg[\delta-3g\zeta -\frac{5}{2}F\eta -\frac{7}{4}F^*\lambda -\frac{3}{4}G \\
&+ 6Fg-\frac{1}{2}Fg^{\star}\lambda+4F^{\star}g\eta + \frac{3}{2}Gg\eta^{\star}-\frac{1}{2}G^{\star}g\lambda -\frac{1}{2}Gg^{\star}\eta\\
&+ 3\mu\Bigg(\frac{3}{2}F\chi+\frac{1}{2}Gg^{\star}\chi-\frac{1}{2}F^{\star}g\chi-3Fg + \frac{1}{2}Gg\chi^{\star}\Bigg)\Bigg]\;,
\label{eq:deltafull}
\end{split}
\end{equation}
\end{subequations}
where $\eta$ and $\lambda$ are the following dimensionless spin-2 and spin-4 quantities, based on 4th-order moments of the light distribution,
\begin{equation}
\begin{split}
\eta &\equiv \frac{1}{\xi S_{0}}\int d^2\theta I(\theta) \theta^3 \theta^{\star} \\
&= \frac{(Q_{1111}-Q_{2222})+2i(Q_{1112}+Q_{1222})}{\xi}\;,\label{eq:eta}
\end{split}
\end{equation}
\begin{equation}
\begin{split}
\lambda &\equiv \frac{1}{\xi S_{0}}\int d^2\theta I(\theta) \theta^4 \\
&= \frac{(Q_{1111}-6Q_{1122}+Q_{2222})+4i(Q_{1112}-Q_{1222})}{\xi}\;,
\end{split}
\end{equation}
and 
\begin{equation}
\mu \equiv \frac{Tr(Q)^2}{\xi} \label{eq:muDef}=\frac{(Q_{11}+Q_{22})^2}{\xi}
\end{equation}
is a dimensionless spin-0 quantity that describes the radial profile of the source. The quantities on the left-hand side, $\zeta^{\mathrm{s}}$ and $\delta^{\mathrm{s}}$, denote the intrinsic, pre-lensing spin-1 and spin-3 distortions of the source. The quantities on the right-hand side, $\zeta$, $\delta$, $\eta$, $\lambda$, $\chi$, denote the observable distortions. Hence, \autoref{eq:flexionfull} relates the intrinsic distortions with apparent, lensing-induced ones $(\zeta, \delta)$ via different combinations of the lensing fields, involving moments up to order four of the light distribution. These equations are conceptually equivalent to the more familiar one for spin-2 distortion relating the intrinsic galaxy ellipticity to the shear and the observed ellipticity \citep[e.g.][]{Seitz97}.  Note that we dropped all moments  higher than fourth order in the derivation of \autoref{eq:flexionfull} as we regard them to be practically immeasurable. Even then, the occurrence of terms like $\eta$ and $\lambda$ is worrying given the limited significance of typical background galaxies. For this reason, \cite{Leonard07,Okura08,Leonard11} have proposed simplified variants of these equations, exploiting the following additional approximations:
\begin{itemize}
\item $\eta$ and $\lambda$ are small and thus negligible;
\item The shear and the convergence are small.
\end{itemize}
Incorporating these simplifications, a linear relation between $\zeta$ and $\delta$ and the flexion fields can be found, which allows us to directly solve for
\begin{subequations}
\begin{align}
F& \simeq \Bigg \langle \frac{\zeta}{(9/4)-3\mu}\Bigg \rangle \ \mathrm{and}\label{eq:simpleF}\\
G &\simeq \frac{4}{3}\langle \delta \rangle\label{eq:simpleG}\;.
\end{align}
\label{eq:simpleflex}
\end{subequations}
We shall investigate the validity of this simplified formula in \autoref{sec:tests}.

\section{Morphological dependence of flexion measurements}
\label{sec:morphology}

In this section we seek insight into the physical meaning and importance of the terms appearing in \autoref{eq:flexionfull}. This will guide our approach to simplify these equations such that we can come up with practically useful, yet accurate flexion estimators. 

\subsection{Galaxy ellipticities $\chi$ and $\eta$}

In \autoref{eq:flexionfull}, the ellipticity of a galaxy is described by two quantites, $\eta$ and $\chi$.  
\citet{Seitz95} showed that $\chi$ transforms under lensing as
\begin{equation}
\chi^{\mathrm{s}}=\frac{\chi-2g+g^2\chi^{\star}}{1+|g|^2-2\Re(g\chi^{\star})}\;.
\label{eq:chi_g}
\end{equation}
It is easy to see that at first order in the shear
\begin{equation}
\langle \chi \rangle \simeq 2g(1 - \langle \chi \chi \rangle)\;, \label{eq:chiresp}
\end{equation}
where the term in brackets is the so-called \emph{responsivity}, and $\langle \chi \chi \rangle$ denotes the variance of the $\chi$-distribution.
Even though $\chi$ is not an unbiased estimator of the shear -- hence the need for the first-order responsivity correction factor -- it is normally used in weak-lensing applications since it has simpler noise properties compared to the unbiased shear estimator
\begin{equation}
\epsilon \equiv \frac{a-b}{a+b}e^{2i\phi}\;,
\end{equation}
which is related to $\chi$ according to
\begin{equation}
\chi=\frac{2\epsilon}{1+|\epsilon|^2}\;. \label{eq:chieps}
\end{equation}

The other quantity describing the ellipticity of the object is $\eta$. It is a spin-2 quantity, defined by a ratio of fourth-order moments of the light distribution. 
Assuming an object with elliptical isophotes, we show in \autoref{sec:eta_eps} that $\eta$ is related to $\epsilon$ by
\begin{equation}
\eta=\frac{3\epsilon(1+|\epsilon|^2)}{1+4|\epsilon|^2+|\epsilon|^4}=\frac{a^4-b^4}{a^4+\frac{2}{3}a^2b^2 +b^4}e^{2i\phi}\;.
\label{eq:etaeps}
\end{equation}
We also compute how $\eta$ transforms under the lens equation and derive its responsivity analogously to the responsivity of $\chi$,
\begin{equation}
\langle \eta \rangle \simeq 3g\Bigg(1-\frac{4}{3}\langle \eta \eta \rangle \Bigg)\;,
\label{eq:etaresp}
\end{equation}
where $\langle \eta \eta \rangle$ is the variance of the $\eta$-distribution. Note that \citet{Okura09} introduced $\eta$ as higher-order shear estimator (Equation 12 in their paper), but did not consider its responsivity. 

\begin{figure}
\includegraphics[width=\linewidth, angle=0]{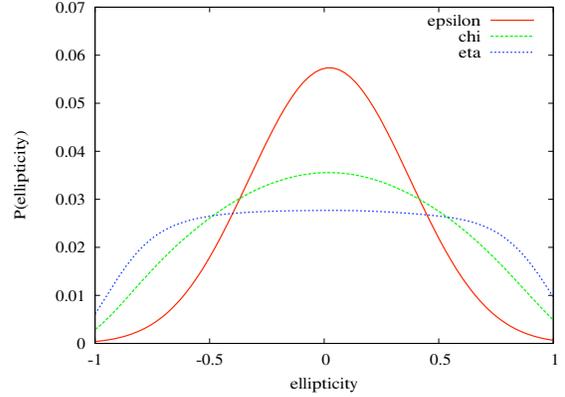}
\caption{Distributions of $\chi_{i}$ (green) and $\eta_{i}$ (blue) under the assumption that $\epsilon_{i}$ (red) has a Gaussian distribution with standard deviation $\sigma_{\epsilon}=0.3$. The relations between $\chi$, $\eta$ and $\epsilon$ are given in \autoref{eq:chieps} and \autoref{eq:etaeps}.} \label{fig:elDis}
\end{figure}

To quantify the responsivity corrections for $\chi$ and $\eta$ we need to compute the dispersions of the respective distributions. This is, however, hampered by their non-linear relation to the proper ellipticity $\epsilon$. Even if we follow the usual assumption that both components of $\epsilon$ have Gaussian distributions, this property is lost for $\chi_i$ and $\eta_i$. In \autoref{fig:elDis} we show what these distributions actually look like under the assumption that $\epsilon_{i}$ has a Gaussian distribution with $\sigma_{\epsilon_i} = 0.3$. 
Calculating the analytic form of the $\chi$- and $\eta$-distributions and their variances is thus not trivial, a numerical integration on the other hand provides us with sufficiently accurate fitting formulae, valid in the range  $\sigma_{\epsilon_i} \in [0.2,0.4]$:
\begin{equation}
\langle \chi \chi \rangle = -1.87\sigma^{2}_{\epsilon}+2.04\sigma_{\epsilon}+0.02, \label{eq:sigmachi}
\end{equation}
and
\begin{equation}
\langle \eta \eta \rangle = -2.04\sigma^2_{\epsilon}+1.96\sigma_{\epsilon}+0.13. \label{eq:sigmaeta}
\end{equation}

\subsection{Galaxy concentration $\mu$}

The quantity $\mu$ defined in \autoref{eq:muDef} is the ratio of two quantities with spin-0, $Tr(Q)^2$ and $\xi$, both describing the size of the object, one in terms of 2nd-order moments, the other in terms of 4th-order moments. This ratio therefore encodes the steepness of the object's surface brightness profile. In other words, $\mu$ can be seen as a concentration parameter: the smaller $\mu$, the more concentrated the object is.
The intrinsic value of $\mu$ depends on the morphology of the galaxy which is used to measure the flexion. It is easy to show analytically that in the case of a top-hat surface brightness profile, $\mu$ equals $\tfrac{3}{4}$. A numerical integration of more realistic radial profiles (e.g. S\'ersic profiles) reveals that $\mu$ typically falls into the range $[0.001,0.5]$ (cf. \autoref{tab:mu}).

\begin{table}
\centering
   \caption{Values of the $\mu$ parameter, as defined in \autoref{eq:muDef} for three different S\'ersic index and three different ellipticities. $n_{s}=0.5$ correspond to a Gaussian profile, while $n_{s}=4.0$ corresponds to a de Vaucouleurs profile.}
\label{tab:mu}
\begin{tabular}{|l|c|c|c|}
  \hline
  &$|\epsilon|=0$&$|\epsilon|=0.3$&$|\epsilon|=0.6$\\
  \hline
  $n_{s}$=0.5&0.5&0.43&0.35\\
  $n_{s}$=2.0&0.1&0.08&0.06\\
  $n_{s}$=4.0&0.005&0.004&0.002\\
  \hline
\end{tabular} 
\end{table}

In general $\mu$ depends not only on the galaxy's S\'ersic index but also on its intrinsic ellipticity. For a fixed S\'ersic index, $\mu$ decreases with increasing ellipticity.
The $\mu$-terms in \autoref{eq:flexionfull} arise from the flexion-induced shift of the centroid position (cf.\ \autoref{sec:flexion_trafo}). More concentrated galaxies are less susceptible to this shift, hence the observable flexion effects are stronger for those. Flexion effects are therefore most easily observable for circular galaxies, but also for such with low $\mu$.

\subsection{Spin-4 distortion $\lambda$}

The remaining quantity $\lambda$ has spin 4 and is thus related to a distortion of the galaxy with fourfold rotational symmetry. We expect the pre-lensing value of $\lambda$ to be small given the fact that square-shaped galaxies appear to be less numerous in Nature \citep{Kayser89}. Moreover, its lensing induced change can only be related to terms like $GF$, $g^2$, $gGF^*$ or any other product of lensing fields corresponding to a spin-4 field. This implies that all terms of \autoref{eq:flexionfull} containing $\lambda$ are of higher order in the lensing fields and are therefore neglected in what follows.

\section{Flexion estimators for stronger lenses}
\label{sec:new_estimators}

As we argued above, the assumptions made at the end of \autoref{sec:flexion_moments}, which led to the linear flexion estimators given in \autoref{eq:simpleflex}, appear to be too restrictive, especially for cluster-lensing applications. In particular, shear and convergence cannot be neglected at distances of $[0.1-1]\,\mathrm{Mpc/}h$ from the cluster centre, where flexion has been measured so far \citep{Leonard07,Okura08}. There the higher-order terms we kept in the derivation of  \autoref{eq:flexionfull} become important. 
At larger scales, where the assumptions of small shear and convergence hold, the signal is very weak due to the rapid decrease of the strengh of flexion as a function of scale, and therefore very difficult to measure. So far, no detection of a flexion signal has been claimed on scales larger than $1.5\,\mathrm{Mpc\,h^{-1}}$.
To first order in the shear, the main corrections to the simple flexion estimator given in \autoref{eq:simpleflex} come from terms of the form (shear $\times$ flexion). 

While the change of spin-1 or spin-3 distortions under lensing given in \autoref{eq:flexionfull} is valid for any single object, a direct solution for the lensing fields is impossible since we do not know the intrinsic values $\zeta^{\mathrm{s}}$ and $\delta^{\mathrm{s}}$.\footnote{For the same reason it is impossible to infer the shear from a measurement of the ellipticity of a single galaxy since its intrinsic ellipticity is unknown.} However, if these distortions obey isotropy, they should average to zero as long as the background source ensemble is not affected by lensing. This means that  the left-hand sides of \autoref{eq:zetafull} and \autoref{eq:deltafull} should vanish in the mean. We proceed to calculate flexion estimators by averaging the right-hand sides of \autoref{eq:zetafull} and \autoref{eq:deltafull} -- assuming that the lensing fields are constant for all galaxies considered in the average --
according to the following scheme:
\begin{enumerate}
\item Taylor-expand the denominator to first order in $g$;
\item Average all terms in the equation obtained this way;
\item Drop all terms of second order in $g$;
\item Use \autoref{eq:chiresp} and \autoref{eq:etaresp} to link $\langle \chi \rangle$ and $\langle \eta \rangle$ to $g$;
\item Drop all the terms containing products between $g$ and $\langle \chi \rangle$ or $\langle \eta \rangle$ appears since they effectively scale as $g^2$;
\item Neglect all $\lambda$-terms as they are of at least second order in the lensing fields;
\item Express $\langle \zeta \rangle$ in terms of $F$ using \autoref{eq:simpleF} and $\langle \delta \rangle$ in terms of $G$ using \autoref{eq:simpleG}.
This approximation is sufficient since any further correction involves terms of second order in $g$;
\item \label{it:abc} Neglect terms like $\langle \mu \eta \Re(g\chi^{\star})\rangle$, or in general terms like $\langle abc \rangle$, where $a$, $b$ and $c$ represent some morphological properties of the galaxy. These terms appear because of the Taylor expansion of the denominators in \autoref{eq:zetafull} and \autoref{eq:deltafull}. Each of these terms can give a correction to the equations above at the per-cent level, but practically their measurement would be extremly noisy.
\end{enumerate}
Following these steps, we can reduce the complicated form of \autoref{eq:flexionfull} to a much more compact one:
\begin{subequations}
\label{eq:flexionnew}
\begin{alignat}{2}
\langle \zeta \rangle \simeq\  &\Bigg(\frac{9}{4}-3\langle \mu \rangle -\frac{5}{4}\langle \mu |\chi|^2\rangle\Bigg)F\nonumber\\
&+\Bigg(\frac{9}{4}-\frac{5}{4}\langle \mu |\chi|^2 \rangle- \langle \mu \rangle \Bigg)Gg^{\star}\nonumber\\
&+\Bigg(\frac{15}{2} -3\langle \mu |\chi|^2 \rangle- 2\langle \mu \rangle\Bigg)F^{\star}g \nonumber\\
&-4\langle \mu \chi \rangle F^{\star} - \frac{1}{2}\langle \mu \chi^{\star} \rangle G
\label{eq:zetanew}\\
\langle \delta \rangle \simeq\  & \frac{3}{4}(1-\langle \mu |\chi|^2 \rangle)G+\Bigg(\frac{33}{4} -\frac{3}{4}\langle \mu |\chi|^2 \rangle\Bigg)gF\nonumber\\
&-\frac{9}{2}\langle \chi \mu \rangle F
\label{eq:deltanew}
\end{alignat}
\end{subequations}

For a given set of lensing parameters $F$, $G$ and $g$, the terms on the right-hand side now only depend on the mean concentration and the ellipticity dispersion of the galaxies used to measured flexion.

Even at linear order in the flexion fields (i.e. if the shear-flexion cross terms are neglected), the ellipticity dispersion of the galaxies affects the flexion estimators. This can be seen looking at the first term of the right-hand side of the equations above. The reason is that the lensing-induced centroid shift is proportional not only to the lensing fields but also to the ellipticity of the object: the higher the ellipticity, the more the centroid is shifted. The correction terms we computed, which do not appear in any previous work, take this effect into account. For a S\'ersic-type galaxy with S\'ersic index $n_{s}=1.5$, we found numerically $\langle \mu|\chi|^2 \rangle \simeq 0.05$.

It is interesting to note how galaxy morphology affects the strength of the measured distortion. In the extreme example of circular top-hat galaxies ($\mu=0.75$), $\langle \zeta \rangle$ becomes almost completely insensitive to $F$-flexion, but remains susceptible to $Gg^{\star}$ and $F^{\star}g$. 

\autoref{eq:flexionnew} does not provide flexion estimators, but describes the average measured spin-1 and spin-3 distortions in terms of lensing fields and galaxy properties. 
In principle, assuming that the shear is known -- e.g. from measurements of $\chi$ -- we could invert \autoref{eq:flexionnew} in order to obtain $F$ and $G$ flexion estimators in terms of $\zeta$, $\delta$, $\mu$ and $\chi$\footnote{Near cluster centres, the estimation of $g$ from galaxy ellipiticities is problematic in itself because shear measurements are affected by the presence of flexion  \citep{Schneider08}.}. 
This inversion, however, is in general not necessary since the comparison with a theoretical model can always be done at the level of measured distortions. For this reason we shall always refer to \emph{distortion estimators} in the following rather than \emph{flexion estimators}.

\section{Tests}
\label{sec:tests}

In this section, we assess the performance of the new distortion estimators presented in the previous section by means of dedicated simulations.
The galaxies in our simulations follow an elliptical  S\'ersic-type profile \citep{Sersic63},
\begin{equation}
I(r)=I_{0}\exp \left[-b_{\mathrm{n_s}}\left(\left(\frac{r}{R_e}\right)^{1/n_{\mathrm{s}}}-1\right)\right],\label{eq:Sersic}
\end{equation}
where $R_e$ denotes the radius containing half of the flux, $n_{\mathrm{s}}$ the S\'ersic index, and $b_{\mathrm{n_s}}$ an integration constant that depends on $n_{\mathrm{s}}$ \citep[cf.][]{Graham05}. This type of profile is identical to a Gaussian for $n_\mathrm{s} = 0.5$ and is steeper in the centre for $n_{\mathrm{s}} > 0.5$. 

\begin{figure}
\includegraphics[width=\linewidth]{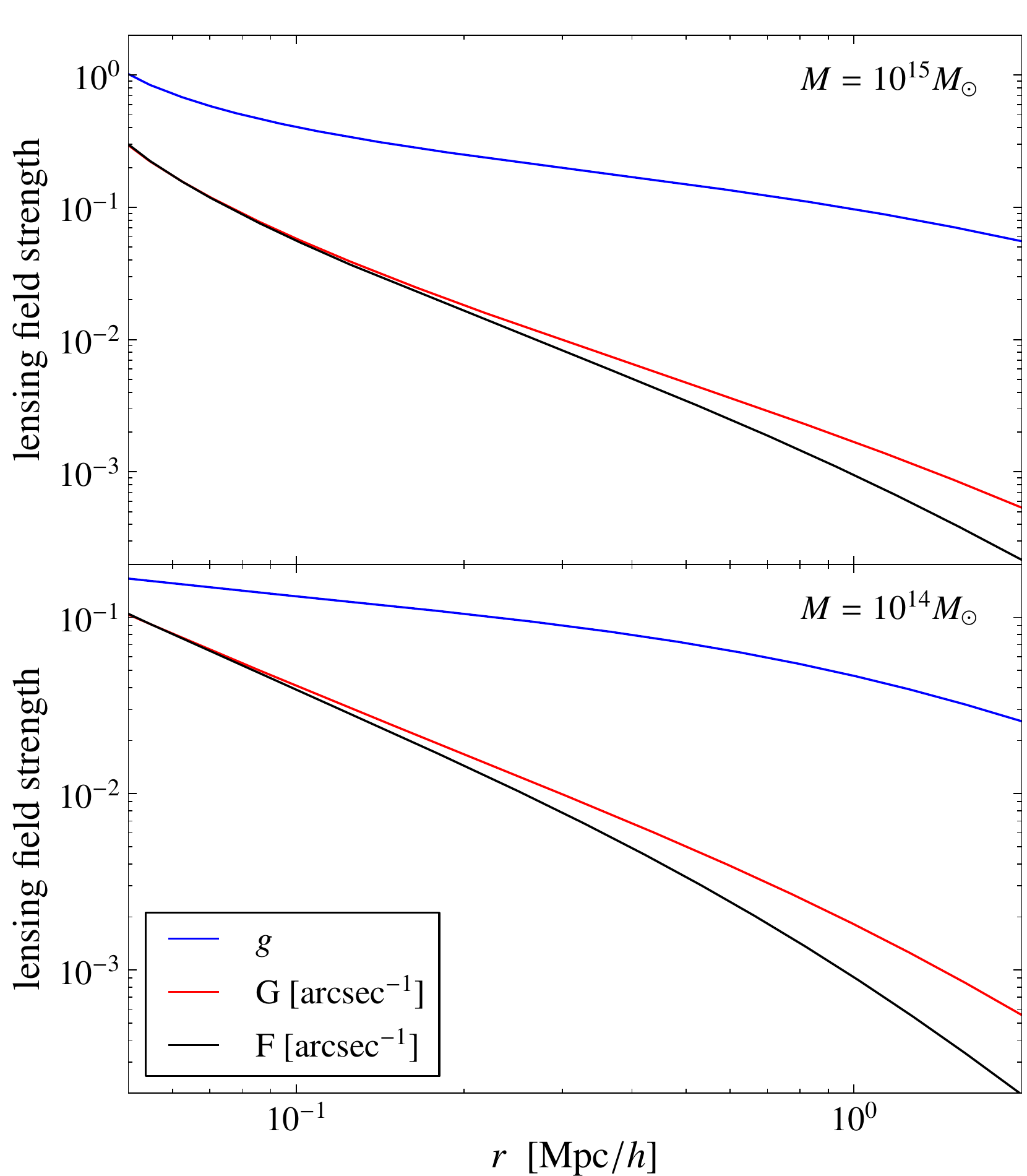}
\caption{Reduced shear $g$ and reduced flexions G and F (top to bottom lines) used in the simulation. The cluster follows an analytic NFW-profile, with masses of $10^{15}\,M_\odot$ (top panel) and $10^{14}\,M_\odot$ (bottom panel) at redshift $z=0.5$. Background galaxies are located at $z_{s}=2$.} 
\label{fig:profile}
\end{figure}

For our test we assumed $n_s=1.5$, which represents the average value for rather bright galaxies in the COSMOS field \citep{Sargent07} and we chose two different radii, $R_e = 2$ pixel and $R_e = 8$ pixel to investigate the size-dependence of flexion estimators. We fixed the pixel-size to be 0.2 arcsec. We lensed the galaxy according to \autoref{eq:lenseqfield}, rotating the reference frame such that the imaginary component of the shear vanishes. The strength of the lensing fields was drawn from analytic NFW profiles \citep{NFW, Bartelmann96, Bacon06} with mass $10^{14}$ or $10^{15}\;M_\odot$ and concentration $c=4.67$ and $c=3.66$, respectively \citep{Neto07}, both at redshift $z=0.5$ with sources at $z_{s}=2.0$. The reduced shear and flexions for this density profile are shown in \autoref{fig:profile}. The crucial point of this setup is that all the relevant lensing fields are obtained from a realistic cluster profile, in particular none of them vanishes in the investigated region. The range of field strengths for the two halos is specified in \autoref{tab:profile}.

\begin{table}
\centering
   \caption{Minimum and maximum strength of the reduced shear and the reduced flexion used in the simulation. These values are drawn from the analytic NFW profiles of \autoref{fig:profile} in the range $r\in[0.15-2] \mathrm{Mpc}/h$.}
\label{tab:profile}
\begin{tabular}{|l|c|c|}
  \hline
  &$M=10^{15} M_\odot$&$M=10^{14} M_\odot$\\
  \hline
  Reduced shear $g$ &0.086-0.30&0.026-0.11\\
  Reduced F [arcsec$^{-1}$]& 0.00065-0.027&0.0001-0.020\\
  Reduced G [arcsec$^{-1}$] &0.0013-0.029&0.0005-0.023\\
  \hline
\end{tabular} 
\end{table}

Before the moments were measured, we recentered the object to account for the lensing-induced centroid shift\footnote{This is a crucial step since the lensing-induced centroid shift has been derived assuming that the centroid is at the coordinate origin of the lens plane.}. We measured the moments of the object and computed $\delta$ and $\zeta$, as well as $\mu$ and the ellipticity $\chi$. In this way we could describe the averaged measured distortions $\langle \zeta \rangle$ and $\langle \delta \rangle$ in terms of the lensing field and the morphological properties of the galaxy sample. We then computed the ratio of the actually measured distortions to the estimated distortions for the two halos, obtained either from \autoref{eq:simpleflex} (linear estimator) or from \autoref{eq:flexionnew} (new estimator derived in this paper).
It is worth noting here that our simulated galaxy images are not affected by noise or convolution with the Point-spread Function of the telescope, thus we can directly measure unweighted moments of the post-lensing galaxy shape.

\begin{figure*}
\begin{minipage}{\linewidth}
\includegraphics[width=\linewidth]{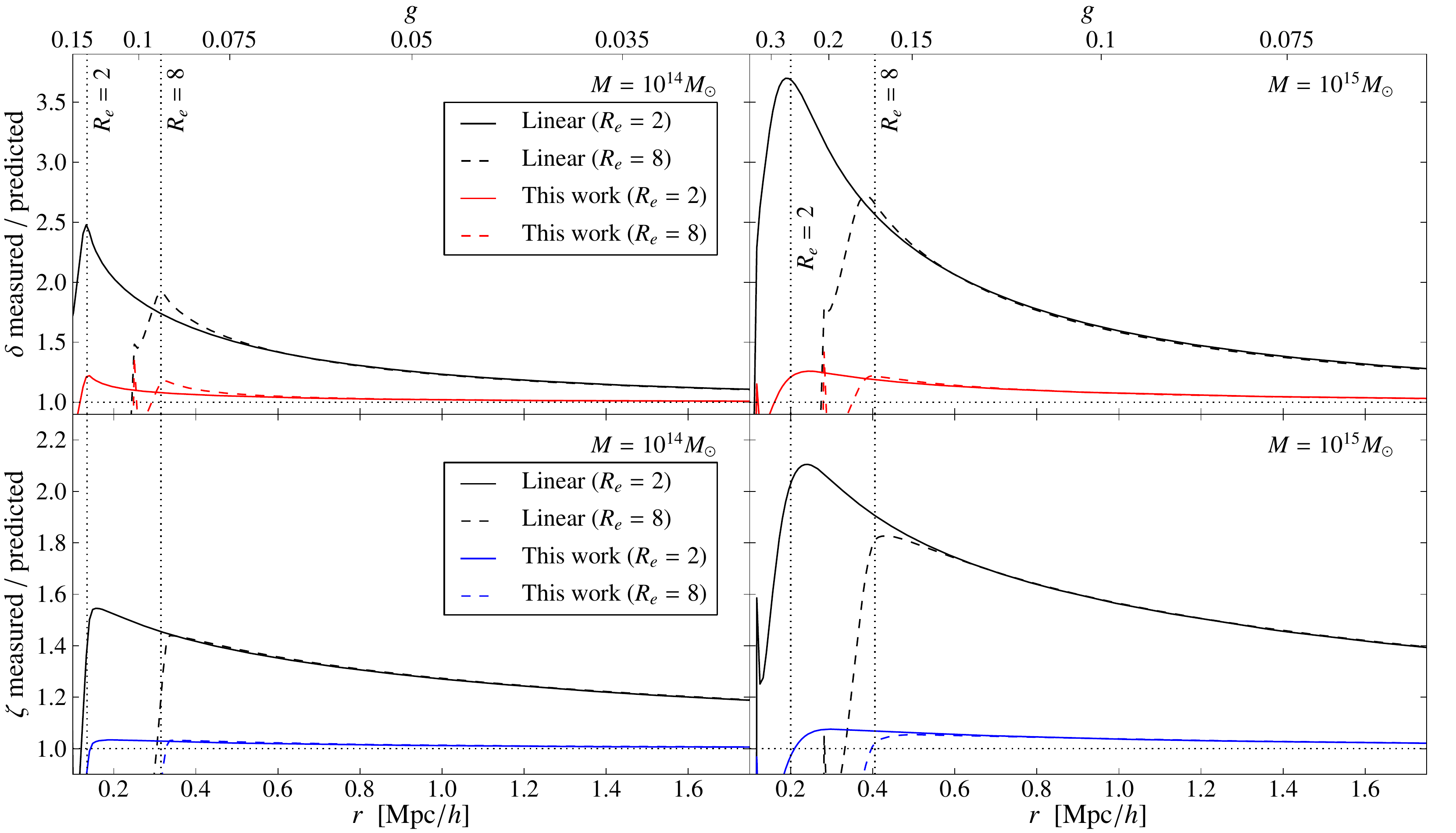}
\caption{Ratio between the measured spin-3 signal $\langle \delta \rangle$ (upper panels) and the spin-1 signal $\langle \zeta \rangle$ (bottom panels), the linear estimator (black) described in \autoref{eq:simpleflex} and the new estimator presented in \autoref{eq:zetanew} (blue) and \autoref{eq:deltanew} (red). The effective radius $R_e$ of the galaxies is set to 2 pixels (solid lines) or 8 pixels (dashed lines). All galaxies were assumed to be intrinsically circular. The two dotted vertical lines show the scale radius where spatially connected multiple images occur (dependent on the galaxy size $R_e$). The left panels refer to the halo of $M=10^{14}\,M_{\odot}$ while the right panels to the $M=10^{15}\,M_{\odot}$ halo. The value of the reduced shear is indicated by the top horizontal axis.} 
\label{fig:flex_nosigma}
\end{minipage}
\end{figure*}

\begin{figure*}
\begin{minipage}{\linewidth}
\includegraphics[width=\linewidth]{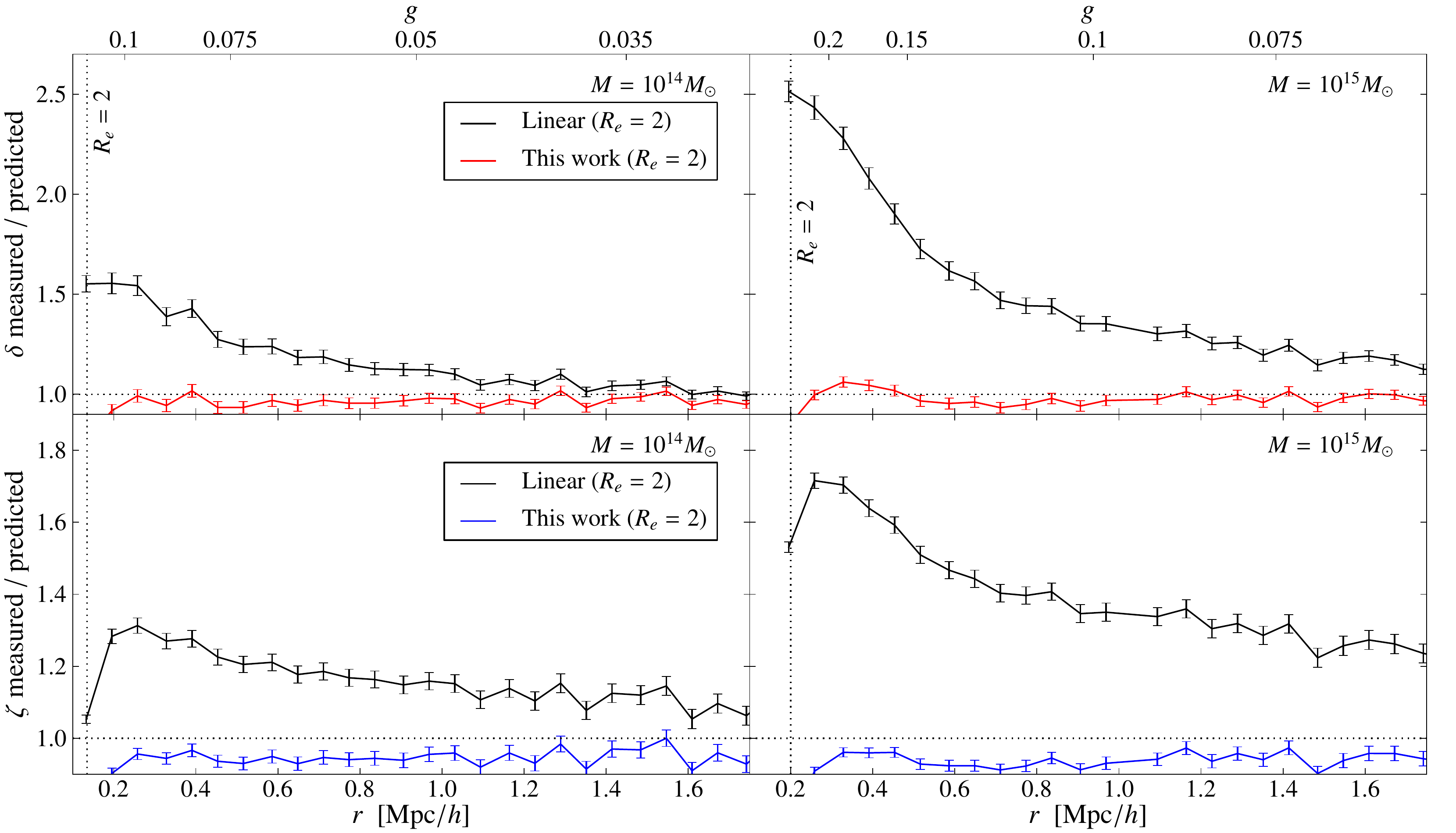}\hfill
\caption{Equivalent to \autoref{fig:flex_nosigma} for galaxies with $R_e=2$ pixel. The average is taken over a galaxy population with ellipticity dispersion $\sigma_{\epsilon}=0.3$. The error bars show the 1-$\sigma$ error of the mean computed by averaging $\zeta$ and $\delta$ over 1000 galaxies.} 
\label{fig:flex_sigma}
\end{minipage}
\end{figure*}

In \autoref{fig:flex_nosigma} and \autoref{fig:flex_sigma}, we show results for circular and realistically elliptical galaxies, respectively. In both cases, zero intrinsic flexion was assumed.
Both figures show that the linear distortion estimators can capture only part of the measured signal, especially at small scales or in general where shear and flexion are stronger. In particular, the difference between the linear estimator and the measured signal can be up to a factor of 3.7 for $\langle \delta \rangle$ or 2.1 for $\langle \zeta \rangle$ for large value of the lensing fields (i.e.\ very close to the centre of a massive cluster) and intrinsically circular sources. If ellipticity dispersion is included in the simulation, the measured signal is up to 2.5 ($\delta$) or 1.7 ($\zeta$) times larger than what the linear estimator predicts.
The reason for this substantial underestimation of the measured signal by the linear estimators is that $\zeta$ and $\delta$ are just a measurement of a spin-1 and a spin-3 field, which can be generated by whatever combination of lensing fields and galaxy morphology measures whose product has either spin-1 or spin-3. Hence, \autoref{eq:zetanew} and \autoref{eq:deltanew}, incorporating the coupling between shear/ellipticity and flexion, provide a much better description of the measured signal, especially if the shear is not very small or the ellipticity dispersion does not vanish. The linear estimator, which generally predicts lower distortions, performs better for intrinsically elliptical objects because lensing-induced distortions are smaller for elliptical rather than circular objects.

However, we notice a residual percent-level bias in our new estimators, which we can explain with the approximations made in deriving \autoref{eq:flexionnew}: Since this bias seems not to depend on the strength of the lensing fields, we must conclude that it originates from neglecting terms accounting for the morphology of the galaxy sample over which the average is taken (cf. \autoref{it:abc} of our derivation scheme). As more such combinations can be found for spin-1 than for spin-3, the estimator for $\zeta$ is more strongly affected than the estimator for $\delta$.

\begin{figure}
\includegraphics[width=\linewidth]{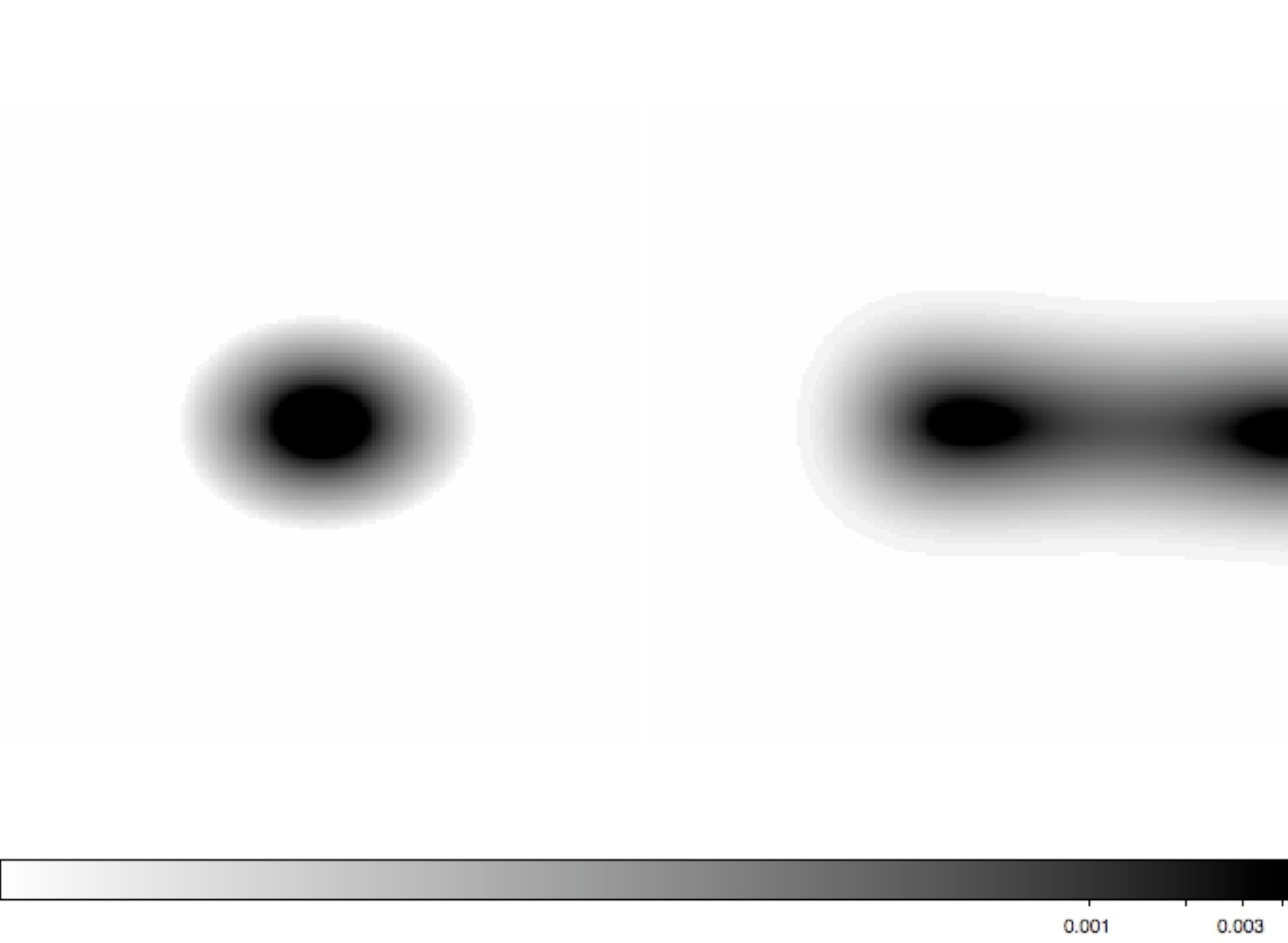}
\caption{Break-down of the flexion concept. \textit{Left panel}: Single image of a circular source with $R_e=2\,\hbox{px}$ lensed by $g=(0.18,0)$, $F=(0.0008,0)$, $G=(0.0011,0)$. \textit{Right panel}: Multiple images of the same source lensed by $g=(0.324,0)$, $F=(0.0067,0)$, $G=(0.007,0)$.}
\label{fig:banana}
\end{figure}

The general conclusion from our tests is that the description of spin-1 and spin-3 distortions in terms of shear, flexion and galaxy morphology that we derived in \autoref{eq:zetanew} and \autoref{eq:deltanew} is valid throughout a wide range of cluster-centric distances. Using them we obtain essentially unbiased estimates of the lensing-induced distortions up to the point, where the very concept of flexion breaks down: when multiple images of an extended source merge \citep{Schneider08}. Such a case is exemplified in \autoref{fig:banana} and the appropriate limits are indicated by dotted vertical lines in \autoref{fig:flex_nosigma} and \autoref{fig:flex_sigma}. Beyond this limit, any flexion measurement is meaningless.

Finally, the bias is also a function of the galaxy size. Since flexion is a dimensional quantity with unit length$^{-1}$, the accuracy of the distortion estimates does not only depend on the strength of the lensing fields but also on the galaxy size: Larger galaxies react more strongly to flexion. Hence, for large objects, the higher-order corrections to \autoref{eq:flexionnew} neglected in this paper become important for lower values of the input flexion field, or farther away from the cluster centre. Also, merging multiple images then occur at larger distances. These effects can be seen in \autoref{fig:flex_nosigma}, where we show results from sources with $R_{e}= 2$ and $R_{e}= 8$ pixels.

\section{Implications}
\label{sec:discussion}

\begin{figure}
\includegraphics[width=\linewidth]{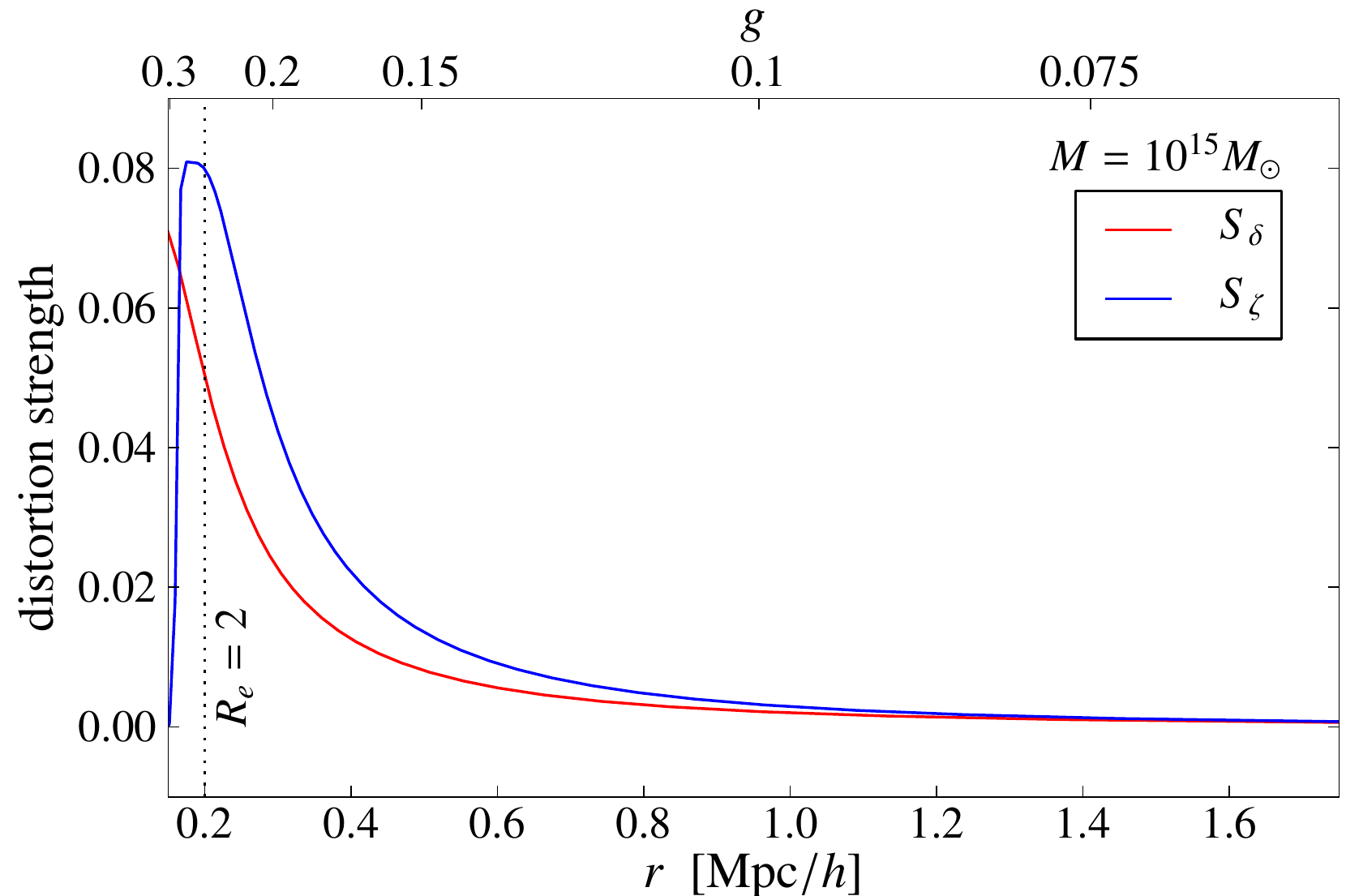}
\caption{Measured $\langle \zeta \rangle$ and $\langle \delta \rangle$ signal as a function of cluster-centric distance for circular galaxies with $R_e=2$ pixels and the halo with $10^{15} M_\odot$. The dotted vertical line represent the radial scale where connected multiple images appear.} \label{fig:comparison}
\end{figure}

Despite several attempts and opposed to the spin-1 flexion distortion, the lensing-induced spin-3 field has not been measured so far \citep{Goldberg05,Leonard07, Okura08, Leonard11}. From a theoretical point of view, this comes as a surprise because the G-flexion field generated by a typical NFW halo is stronger than the F-flexion field (cf.\ \autoref{fig:profile}).
However, when we compare the strength of the measured $\langle \zeta \rangle$ and $\langle \delta \rangle$ signals in \autoref{fig:comparison}, we see that in the radial range $r \in[0.2, 1.0]\,\mathrm{Mpc}/h$, the estimator $\langle \zeta \rangle$ is between 20 \% and 50 \% larger than $\langle \delta \rangle$, hence opposite to the behavior of the F and G fields.

Furthermore, assuming that the noise on the measured moments is Gaussian, we expect the ratio between signal-to-noise ratios of $\zeta$ and $\delta$ to scale as
\begin{equation}
\frac{(S/N)_{\zeta}}{(S/N)_{\delta}}=\sqrt{5}\ \frac{S_{\zeta}}{S_{\delta}},
\end{equation}
where $S_{i}$ denotes the strength of the measured spin-1 ($\zeta$) and spin-3 ($\delta$) field, as shown in \autoref{fig:comparison}.
The pre-factor originates from the definitions of $\zeta$ and $\delta$ in \autoref{eq:zetaDef}. This implies that for a given noise level, the $S/N$ for $\zeta$ can be up to 4 times the one for $\delta$. This simple argument could be one explanation why no spin-3 field around otherwise prominent lenses has been detected yet.

On the other hand, the scale where the lensing fields have a significant amplitude does not only depend on properties of the lens, but also on the dispersion of the intrinsic distortions $\zeta^\mathrm{s}$ and  $\delta^\mathrm{s}$ (as well as $\epsilon^\mathrm{s}$). Since three-fold rotational symmetry is less common in galaxies than skewness, we expect $\sigma_{\delta^\mathrm{s}} < \sigma_{\zeta^\mathrm{s}}$. 
Moreover, as pointed out by \cite{Velander11}, ``light leaking'' from bright objects can also contribute to the measured spin-1 distortion of the object but hardly to the measured spin-3 distortion, possibly causing a bias in the conversion between $\zeta$ and the lensing fields. Thus, despite the lower $S/N$, the lower intrinsic dispersion and the difficulty to generate spurious spin-3 distortions could render the $\delta$ signal more accessible and cleaner than $\zeta$. 

Another interesting point we want to discuss here concerns model-fitting techniques for estimating the strength of the flexions. The standard approach is to model a circular galaxy with a given radial profile, to apply spin-1, spin-2, and spin-3 transformations, and to compare this distorted model with the actual image in order to determine the amplitude of the transformation \citep[e.g.][]{Velander11}.
As we have shown here, a spin-1 and a spin-3 distortion cannot be simply related to flexion, thus the interpretation of the amplitude of the best-fitting transformation in terms of lensing fields is not straightforward. While shearing an elliptical galaxy is equivalent to applying a magnification, a rotation and a pure spin-2 distortion to a circular galaxy \citep[in the limit of small shear:][]{Kuijken06}, the same is not true for the flexions, even if $F$, $G$ and $g$ are small. The reason for this somewhat counterintuitive behavior stems from the dependence of the response to flexion on the galaxy ellipticity (cf. \autoref{eq:zetanew} and \autoref{eq:deltanew}). In other words, shearing and flexing an elliptical galaxy is in general \emph{not} the same as applying magnification, rotation, spin-1, spin-2 and spin-3 distortions to a circular object, just because the amount spin-2 distortion differs in the two cases.

\begin{figure}
\includegraphics[width=\linewidth]{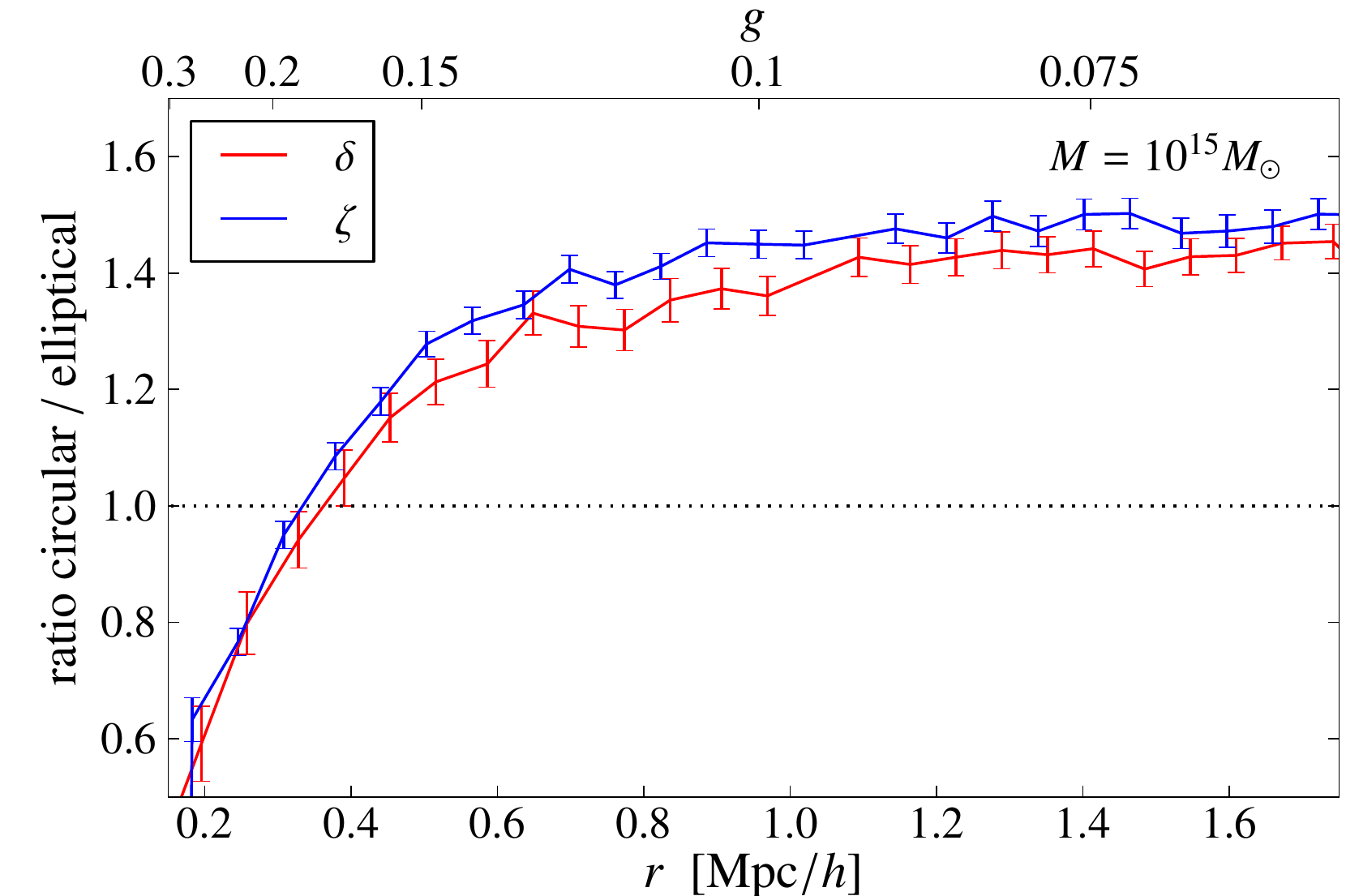}
\caption{Ratio between the average spin-1 and spin-3 distortions for intrinsically circular and elliptical S\'ersic-type sources. For the circular galaxies, the spin-2 distortions was adjusted such as to generate the entire post-lensing ellipticity. Plotted are average and 1-$\sigma$ errors of 1000 galaxies with ellipticities sampled from a Gaussian distribution with $\sigma_{\epsilon}=0.3$.}
\label{fig:modelfitting}
\end{figure}
In \autoref{fig:modelfitting}, we compare the strength of the average spin-1 and spin-3 distortions of intrinsically circular or elliptical galaxies. For the former, the effective shear was chosen such as to fully account for the apparent post-lensing ellipticity (thus combining intrinsic and actual lensing-induced ellipticity) of the galaxies used in the latter case.
At large radii, employing a circular source model leads to a 40\% overestimation of the true spin-1 and spin-3 distortions. In contrast, at small radii, the actual signal would rather be underestimated. The complicated impact of the assumption of circular source models on the flexion-distortion estimates would thus require the inclusion of the intrinsic source ellipticity into the set of model/lensing parameters. Besides increasing the numerical demands of doing so, the strong degeneracy between intrinsic and apparent post-lensing ellipticities, however, impedes such an approach without additional assumptions on the strength of the shear.

\section{Conclusions}
\label{sec:conclusions}

In this work we show that measurements of spin-1 ($\zeta$) and spin-3 distortions ($\delta$) of galaxy shapes in terms of surface-brightness moments cannot be related in a straightforward way to the $F$- and $G$ flexions caused by gravitational lensing, in particular not for strongly elliptical galaxies or near the centers of galaxy clusters. In the latter case, the reduced shear is not small, thus terms coupling shear and flexion produce non-negligible spin-1 and spin-3 distortions, which contribute to the distortions induced by the flexion alone. A similar effect applies in cases where the observed ellipticity is not small. These coupling terms imply that in general the measured signal will typically be larger than the naively expected flexion signal.

Furthermore, lensing causes a centroid shift that is approximately linear in the flexion fields and it depends on the ellipticity and the concentration of the source galaxy. Hence, when flexion-related distortions measured from individual objects are averaged over a source galaxy population with given distributions of ellipticity and concentration, the measured signal will depend not only on the strength of the lensing fields but also on average galaxy properties.

Our new derivation of the lensing-induced changes of the spin-1 and spin-3 distortion estimators accounts for all of the effects mentioned above. This work extends that of \cite{Okura08} -- yielding a linear relation between flexion and spin-1 and spin-3 distortions (cf. \autoref{eq:simpleflex}) -- in two ways:
\begin{itemize}
\item We include shear-flexion coupling terms in the transformation of the flexion estimators $\zeta$ and $\delta$;
\item We compute how the average of $\delta$ and $\zeta$ over a population of galaxies is affected by the galaxy ellipticity dispersion.
\end{itemize}

We provide two more detailed equations relating $\langle \zeta \rangle$ and $\langle \delta \rangle$ with the reduced shear and the reduced flexion. These equations take the contribution to the measured signal, originating from concentration and ellipticity dispersions of the galaxy population, explicitly into account. Treating the dependence on galaxy morphology is important, since it impacts on the distortion estimates even if shear-flexion cross-terms can be neglected.

By means of several numerical tests, we show that simple linear estimators can lead to biases of up to 85\% and 150\% for $\zeta$ and $\delta$, respectively. This bias depends in general on the strength of the lensing field, on the intrinsic ellipticity dispersion and on the concentration of the galaxies. Moreover, we showed that the description of the spin-1 and spin-3 distortions that we derived in this paper is valid (with a bias $\leq 7\%$ (spin-1 distortion) and $\leq 3\%$  (spin-3 distortion)) even close to the point where the flexion concept breaks down due to the merging of multiple images.

We finally find that the measurement of the spin-1 distortion is always stronger (at scales where flexion is usually measured) than the spin-3 distortion. This, combined with the different noise properties of the $\zeta$ and $\delta$ estimators, could be an explanation of why a spin-3 field has not been detected so far.

\section*{Acknowledgements}
We would like to thank Carlo Giocoli for providing convergence, shear and flexion maps used in the tests presented in \autoref{sec:tests}. Moreover we would like to thank the anonymous referee for his/her punctual comments to the original manuscript and for his/her valuable suggestions.
This work was supported in parts by the EU-RTN ``DUEL'', the Heidelberg Graduate School of Fundamental Physics, the IMPRS for Astronomy \& Cosmic Physics at the University of Heidelberg and the Transregio-Sonderforschungsbereich TR-33 of the Deutsche Forschungsgemeinschaft. MV is supported by a STFC Rolling Grant RA0888. PM is supported by the U.S.\ Department of Energy under Contract No.\ DE- FG02-91ER40690.
\bibliographystyle{mn2e}
\bibliography{biblio}

\begin{thebibliography}{}

\bibitem[\protect\citeauthoryear{{Bacon}, {Goldberg}, {Rowe} \&
  {Taylor}}{{Bacon} et~al.}{2006}]{Bacon06}
{Bacon} D.~J.,  {Goldberg} D.~M.,  {Rowe} B.~T.~P.,    {Taylor} A.~N.,  2006,
  \mnras, 365, 414

\bibitem[\protect\citeauthoryear{{Bartelmann}}{{Bartelmann}}{1996}]{Bartelmann96}
{Bartelmann} M.,  1996, \aap, 313, 697

\bibitem[\protect\citeauthoryear{{Bartelmann} \& {Schneider}}{{Bartelmann} \&
  {Schneider}}{2001}]{Bartelmann01}
{Bartelmann} M.,  {Schneider} P.,  2001, \physrep, 340, 291

\bibitem[\protect\citeauthoryear{{Goldberg} \& {Bacon}}{{Goldberg} \&
  {Bacon}}{2005}]{Goldberg05}
{Goldberg} D.~M.,  {Bacon} D.~J.,  2005, \apj, 619, 741

\bibitem[\protect\citeauthoryear{{Graham} \& {Driver}}{{Graham} \&
  {Driver}}{2005}]{Graham05}
{Graham} A.~W.,  {Driver} S.~P.,  2005, \pasa, 22, 118

\bibitem[\protect\citeauthoryear{{Hawken} \& {Bridle}}{{Hawken} \&
  {Bridle}}{2009}]{Hawken09}
{Hawken} A.~J.,  {Bridle} S.~L.,  2009, \mnras, 400, 1132

\bibitem[\protect\citeauthoryear{{Irwin}, {Shmakova} \& {Anderson}}{{Irwin}
  et~al.}{2007}]{Irwin07}
{Irwin} J.,  {Shmakova} M.,    {Anderson} J.,  2007, \apj, 671, 1182

\bibitem[\protect\citeauthoryear{{Kayser}, {Refsdal}, {Weiss} \&
  {Schneider}}{{Kayser} et~al.}{1989}]{Kayser89}
{Kayser} R.,  {Refsdal} S.,  {Weiss} A.,    {Schneider} P.,  1989, \aap, 214, 4

\bibitem[\protect\citeauthoryear{{Kuijken}}{{Kuijken}}{2006}]{Kuijken06}
{Kuijken} K.,  2006, \aap, 456, 827

\bibitem[\protect\citeauthoryear{{Leonard}, {Goldberg}, {Haaga} \&
  {Massey}}{{Leonard} et~al.}{2007}]{Leonard07}
{Leonard} A.,  {Goldberg} D.~M.,  {Haaga} J.~L.,    {Massey} R.,  2007, \apj,
  666, 51

\bibitem[\protect\citeauthoryear{{Leonard}, {King} \& {Goldberg}}{{Leonard}
  et~al.}{2011}]{Leonard11}
{Leonard} A.,  {King} L.~J.,    {Goldberg} D.~M.,  2011, \mnras, 413, 789

\bibitem[\protect\citeauthoryear{{Navarro}, {Frenk} \& {White}}{{Navarro}
  et~al.}{1997}]{NFW}
{Navarro} J.~F.,  {Frenk} C.~S.,    {White} S.~D.~M.,  1997, \apj, 490, 493

\bibitem[\protect\citeauthoryear{{Neto}, {Gao}, {Bett}, {Cole}, {Navarro},
  {Frenk}, {White}, {Springel} \& {Jenkins}}{{Neto} et~al.}{2007}]{Neto07}
{Neto} A.~F.,  {Gao} L.,  {Bett} P.,  {Cole} S.,  {Navarro} J.~F.,  {Frenk}
  C.~S.,  {White} S.~D.~M.,  {Springel} V.,    {Jenkins} A.,  2007, \mnras,
  381, 1450

\bibitem[\protect\citeauthoryear{{Newman} \& {Penrose}}{{Newman} \&
  {Penrose}}{1962}]{Newman62}
{Newman} E.,  {Penrose} R.,  1962, Journal of Mathematical Physics, 3, 566

\bibitem[\protect\citeauthoryear{{Okura} \& {Futamase}}{{Okura} \&
  {Futamase}}{2009}]{Okura09}
{Okura} Y.,  {Futamase} T.,  2009, \apj, 699, 143

\bibitem[\protect\citeauthoryear{{Okura}, {Umetsu} \& {Futamase}}{{Okura}
  et~al.}{2007}]{Okura07}
{Okura} Y.,  {Umetsu} K.,    {Futamase} T.,  2007, \apj, 660, 995

\bibitem[\protect\citeauthoryear{{Okura}, {Umetsu} \& {Futamase}}{{Okura}
  et~al.}{2008}]{Okura08}
{Okura} Y.,  {Umetsu} K.,    {Futamase} T.,  2008, \apj, 680, 1

\bibitem[\protect\citeauthoryear{{Sargent}, {Carollo}, {Lilly}, {Scarlata},
  {Feldmann}, {Kampczyk}, {Koekemoer}, {Scoville}, {Kneib} \& {et
  al.}}{{Sargent} et~al.}{2007}]{Sargent07}
{Sargent} M.~T.,  {Carollo} C.~M.,  {Lilly} S.~J.,  {Scarlata} C.,  {Feldmann}
  R.,  {Kampczyk} P.,  {Koekemoer} A.~M.,  {Scoville} N.,  {Kneib} J.,    {et
  al.} 2007, \apjs, 172, 434

\bibitem[\protect\citeauthoryear{{Sch\"{a}fer}, {Heisenberg}, {Fotios
  Kalovidouris} \& {Bacon}}{{Sch\"{a}fer} et~al.}{2011}]{Schaefer11}
{Sch\"{a}fer} B.~M.,  {Heisenberg} L.,  {Fotios Kalovidouris} A.,    {Bacon}
  D.~J.,  2011, ArXiv e-prints

\bibitem[\protect\citeauthoryear{{Schneider} \& {Er}}{{Schneider} \&
  {Er}}{2008}]{Schneider08}
{Schneider} P.,  {Er} X.,  2008, \aap, 485, 363

\bibitem[\protect\citeauthoryear{{Seitz} \& {Schneider}}{{Seitz} \&
  {Schneider}}{1995}]{Seitz95}
{Seitz} C.,  {Schneider} P.,  1995, \aap, 297, 287

\bibitem[\protect\citeauthoryear{{Seitz} \& {Schneider}}{{Seitz} \&
  {Schneider}}{1997}]{Seitz97}
{Seitz} C.,  {Schneider} P.,  1997, \aap, 318, 687

\bibitem[\protect\citeauthoryear{{S{\'e}rsic}}{{S{\'e}rsic}}{1963}]{Sersic63}
{S{\'e}rsic} J.~L.,  1963, Boletin de la Asociacion Argentina de Astronomia La
  Plata Argentina, 6, 41

\bibitem[\protect\citeauthoryear{{Velander}, {Kuijken} \&
  {Schrabback}}{{Velander} et~al.}{2011}]{Velander11}
{Velander} M.,  {Kuijken} K.,    {Schrabback} T.,  2011, \mnras, 412, 2665

\bibitem[\protect\citeauthoryear{{White} \& {Rees}}{{White} \&
  {Rees}}{1978}]{White78}
{White} S.~D.~M.,  {Rees} M.~J.,  1978, \mnras, 183, 341

\end{thebibliography}

\appendix
\onecolumn

\section{Transformation of $\zeta$ and $\delta$}
\label{sec:flexion_trafo}

Using the definition of $\zeta$ from \autoref{eq:zetaDef} and the second-order lens equation (\autoref{eq:lenseqfield}), we can relate the lensing fields with the spin-1 distortion on the source plane
\begin{equation}
\begin{split}
\zeta^{s}&=\frac{\int d^2 \beta I^{s}(\beta) \beta^2\beta^{\star}}{\int d^2 \beta I^{s}(\beta) (\beta \beta^{\star})^2}
 \simeq \frac{(1-\kappa)^5}{\xi^{s}} \int d^2\theta I(\theta)\det A \Bigg(\theta-g\theta^{\star}-\frac{1}{4} F^{\star}\theta^2-\frac{1}{2}F\theta \theta^{\star} -\frac{1}{4}G (\theta^{\star})^2\Bigg)^2\Bigg(\theta^{\star}-g^{\star}\theta-\frac{1}{4} F(\theta^{\star})^2-\frac{1}{2}F^{\star}\theta \theta^{\star} -\frac{1}{4}G^{\star} \theta^2\Bigg) \simeq \\
&\simeq \frac{1}{\xi^{s}}\int d^2 \theta I(\theta)\Bigg(\theta^2\theta^{\star}-g^{\star}\theta^{3}+\frac{3}{2}F^{\star}g^{\star}\theta^4 -\frac{1}{4}G^{\star}\theta^4 -2F^{\star}\theta^3\theta^{\star}+\frac{3}{2}Fg^{\star}\theta^3\theta^{\star}-2g\theta(\theta^{\star})^2-\frac{9}{4}F(\theta \theta^{\star})^2+3F^{\star}g(\theta \theta^{\star})^2+\frac{7}{2}Fg\theta (\theta^{\star})^3\\
&-\frac{1}{2}gG(\theta^{\star})^4\Bigg)\;.
\label{eq:zetacalc}
\end{split}
\end{equation}
In the same way we compute how $\delta$ transforms:
\begin{equation}
\begin{split}
\delta^{s}&=\frac{\int d^2 \beta I^{s}(\beta) \beta^3}{\int d^2 \beta I^{s}(\beta) (\beta \beta^{\star})^2}
 \simeq \frac{(1-\kappa)^5}{\xi^{s}} \int d^2\theta I(\theta)\det A \Bigg(\theta-g\theta^{\star}-\frac{1}{4} F^{\star}\theta^2-\frac{1}{2}F\theta \theta^{\star} -\frac{1}{4}G (\theta^{\star})^2\Bigg)^3 \simeq \\
&\simeq \frac{1}{\xi^{s}} \int d^2 \theta I(\theta) \Bigg(\theta^3-3g \theta^2 \theta^{\star} -\frac{5}{2}F\theta^3\theta^{\star}-\frac{7}{4}F^{\star}\theta^4 -\frac{3}{4}G(\theta \theta^{\star})^2+ 6Fg(\theta \theta^{\star})^2-\frac{1}{2}Fg^{\star}\theta^4+4F^{\star}g\theta^3 \theta^{\star} + \frac{3}{2}Gg(\theta^{\star})^3\theta \\
&-\frac{1}{2}Gg^{\star}\theta^3 \theta^{\star}\Bigg)\;,
\label{eq:deltacalc}
\end{split}
\end{equation}
where $\det A$ is defined in \autoref{eq:determinant}, and
\begin{equation}
\begin{split}
\xi^{s}&= \int d^2\theta I(\theta)\det A \Bigg(\theta-g\theta^{\star}-\frac{1}{4} F^{\star}\theta^2-\frac{1}{2}F\theta \theta^{\star} -\frac{1}{4}G (\theta^{\star})^2\Bigg)^2 \Bigg(\theta^{\star}-g^{\star}\theta-\frac{1}{4} F (\theta^{\star})^2-\frac{1}{2}F^{\star}\theta^{\star} \theta -\frac{1}{4}G^{\star} (\theta)^2\Bigg)^2 \\
&\simeq (1-\kappa)^6 \int d^2\theta I(\theta)\Bigg[(\theta \theta^{\star})^2-2g^{\star}\theta^3\theta^{\star}-2g\theta (\theta^{\star})^3\Bigg] = (1-\kappa)^6 \xi (1-4\Re(g\eta^{\star}))\;.
\end{split}
\end{equation}
In this derivation, we neglected all moments of higher than fourth order since they are practically immesurable. The combination of these two equations leads to \autoref{eq:zetafull} and \autoref{eq:deltafull} without all the terms containg $\mu$ which come from the inclusion of the centroid shift, which is calculated in the following.

\subsection*{Centroid shift}

Normally moments of the light distribution are computed with respect to the centre of light. This means mathematically:
\begin{equation}
\bar{\theta}=\frac{1}{S_{0}}\int d^2 \theta \theta I(\theta)=0\;.
\end{equation}
However, the centroid is not mapped to the origin of the source plane as we assumed in \autoref{eq:zetacalc} and \autoref{eq:deltacalc}. We derive here the mapping between the centroid in the source plane and in the lens plane up to first order in lensing fields:
\begin{equation}
\begin{split}
&\bar{\beta} = \frac{1}{S_{0}}\int d^2 \beta \beta I^{s}(\beta)\simeq\frac{(1-\kappa)^3}{S (1-\kappa)^2}\int d^2\theta I(\theta)\Bigg(\theta-g \theta^{\star}-\frac{1}{4}F^{\star}\theta^{2}-\frac{1}{2}F\theta \theta^{\star}-\frac{1}{4}G(\theta^{\star})^2\Bigg)\Bigg(1-\theta \Bigg(F^{\star}+\frac{g^{\star}F+gG^{\star}}{2}\Bigg)-\\
&\theta^{\star}\Bigg(F+\frac{g^{\star}G+gF^{\star}}{2}\Bigg)\Bigg) \simeq (Tr(Q))(1-\kappa)\Bigg(-\frac{5}{4}F^{\star}\chi -\frac{1}{2}Fg^{\star}\chi -\frac{1}{2}gG^{\star}\chi-\frac{3}{2}F -\frac{1}{2}Gg^{\star}+\frac{1}{2}F^{\star}g+gF\chi^{\star}-\frac{1}{4}G\chi^{\star}\Bigg)
\end{split}
\end{equation}
Using the form of the centroid shift calculated above we can compute:
\begin{equation}
\zeta^{t}=\frac{1}{S_{0}\xi^{s}}\int d^2\beta (\beta-\bar{\beta})^2(\beta^{\star}-\bar{\beta}^{\star})I^{s}(\beta) \simeq \frac{1}{S_{0}\xi^{s}}\Bigg[\int d^2\beta I^{s}(\beta)\beta^2\beta^{\star}-\bar{\beta}^{\star}\int d^2 \beta I^{s}(\beta)\beta^2 -2\bar{\beta} \int d^2\beta I^{s}(\beta)\beta \beta^{\star} \Bigg]
\end{equation}
and 
\begin{equation}
\delta^{t}=\frac{\int d^2\beta (\beta-\bar{\beta})^3 I^{s}(\beta)}{\int d^2 \beta I^{s}(\beta)((\beta-\bar{\beta}) (\beta^{\star}-\bar{\beta}^{\star}))^2} \simeq \frac{1}{\xi(1-\kappa)^6(1-4\Re(g\eta^{\star})}\Bigg[\int d^2\beta I^{s}(\beta)\beta^3-3\bar{\beta}\int d^2 \beta I^{s}(\beta)\beta^2\Bigg]
\end{equation}
where we made use of the fact that to the order we are interested in, $\xi^{s}$ is unaffected by the centroid shift. The first integral in both expressions can be replaced with the expressions computed in \autoref{eq:zetacalc} and \autoref{eq:deltacalc}, while the second and the third integral, including the corrections due to the centroid shift, can be replaced by:
\begin{equation}
\int d^2 \beta I^{s}(\beta)\beta^2 \simeq (1-\kappa)^4 \int d^2 \theta I(\theta) (\theta^2-2g \theta \theta^{\star}) \simeq (1-\kappa)^4(TrQ)(\chi-2g)
\end{equation}
\begin{equation}
\int d^2 \beta I^{s}(\beta)\beta \beta^{\star} \simeq (1-\kappa)^4(TrQ)(1-2\Re(g\chi^{\star})
\end{equation}
since all terms proportional to flexion or $g^2$ will be dropped anyhow due to the multiplication with $\bar{\beta}$. Then we have:
\begin{equation}
\zeta^{t}=\zeta^{s} - \frac{Tr(Q)}{\xi(1-4\Re(g\eta^{\star}))(1-\kappa)^2}\Bigg[(\chi-2g)\bar{\beta}^{\star} + 2(1-2\Re(g\chi^{\star}))\bar{\beta} \Bigg] \label{eq:csZeta}
\end{equation}
and 
\begin{equation}
\delta^{t} \simeq \delta^{s} - \frac{Tr(Q)}{\xi(1-4\Re(g\eta^{\star}))(1-\kappa)^2}\Bigg[3(\chi\bar{\beta})- 6g\bar{\beta}\Bigg]\label{eq:csDelta}
\end{equation}
which lead to \autoref{eq:zetafull} and \autoref{eq:deltafull}, where the upper index "t" has been dropped.

\section{Higher-order source ellipticity}
\label{sec:eta_eps}

We derive here the relation between $\epsilon$ ellipticity and $\eta$ ellipticity for a source with elliptical isophotes. Without loss of generality we assume the source profile to be described by an elliptical gaussian with ellipticity $\epsilon$. Furthermore we select the reference frame such that $\epsilon_2 = 0$. We can then derive this general relation for the moments of order $i+j$:
\begin{equation}
Q^{(i+j)}=\int dx dy x^{i}y^{j} \exp\Bigg(-\frac{1}{2}(1-\epsilon)^2x^2\Bigg) \exp\Bigg(-\frac{1}{2}(1+\epsilon)^2y^2\Bigg)=\frac{\Gamma(\frac{i+1}{2})\Gamma(\frac{j+1}{2})}{(1-\epsilon)^{i+1}(1+\epsilon)^{j+1}}
\end{equation}
Using the expression for $\eta$ in terms of 4th order moments given in \autoref{eq:eta}, we find:
\begin{equation}
\eta = \frac{3\epsilon(1+\epsilon^2)}{1+4\epsilon^2 + \epsilon^4}
\end{equation}

Moreover, we are interested in computing the responsivity of $\eta$. For doing so, it suffices to compute the transformation of $\eta$ under lensing up to first order in the shear. After some algebra it is possible to show that the following relation holds:
\begin{equation}
\eta^{s}\simeq \frac{\eta-3g}{1-4\Re(g\eta^{\star})}
\end{equation}
and therefore
\begin{equation}
\langle \eta \rangle \simeq 3g\Bigg(1-\frac{4}{3}\langle \eta \eta \rangle \Bigg)
\end{equation}

\label{lastpage}
\end{document}